\documentstyle[prl,aps]{revtex}

\def\pq{p_{\rm q}}
\def\pc{p_{\rm c}}
\def\mathrm#1{{\rm #1}}
\def\del{{\rm d}}

\def\del{{\rm d}}
\def\beq{\begin{equation}}
\def\eeq{\end{equation}}
\def\sqr#1#2{{\vcenter{\vbox{\hrule height.#2pt
                       \hbox{\vrule width.#2pt height#1pt \kern#1pt
                       \vrule width.#2pt}
                     \hrule height.#2pt}}}}

\begin{document}
\twocolumn
\title{Three-Dimensional Quantum Percolation Studied 
by Level Statistics}
\author{Atsushi {\sc Kaneko} and Tomi {\sc Ohtsuki}}
\address{Department of Physics, Sophia University, Kioi-cho 7-1, Chiyoda-ku, 
Tokyo 102-8554.}

\maketitle

\begin{abstract}
{
Three-dimensional quantum percolation problems are studied by
analyzing energy level statistics of electrons on 
maximally connected 
percolating clusters.
The quantum percolation threshold $\pq$, which is larger than
the classical percolation threshold $\pc$, becomes smaller when
magnetic fields are applied, i.e., $\pq(B=0)>\pq(B\ne 0)>\pc$.
The critical exponents are found to be consistent with the
recently obtained values of the Anderson model, supporting the
conjecture that the quantum percolation is classified onto the
same universality classes of the Anderson transition.
Novel critical level statistics at the percolation threshold is
also reported.
}
\end{abstract}

\sloppy
\vspace{0.5cm}

%
%
Studies of percolation\cite{stauffer} have been attracting much
interest and a number of applications have been proposed
due to its simplicity.
One of its important applications is to the
transport properties in three-dimensional (3D) solids.
However, at low temperature, the classical percolation picture
is insufficient and the quantum effects should
be seriously taken into account;
thus localization due to the quantum interference effect
vs. percolation becomes a very interesting problem \cite{shapiro}.

Even if a maximum cluster percolates through
the system, the wave function may be localized and
be unable to carry current.
Thus the quantum percolation
threshold $\pq$ is generally larger than the geometrical classical
percolation threshold $\pc$.
At $\pq$, the localization length diverges and the wave function
extends through the percolating cluster.
It is often debated whether this transition is the same as the
Anderson transition \cite{anderson}.
If this is true,
the critical exponent $\nu$ describing the divergence of the
localization length in the quantum percolation problem
would be the same as that in the Anderson transition, and
the scaling theory of localization would be valid \cite{gang,kawabata}.
Then, a number of studies on Anderson transition\cite{LR,KM} can be
used to study quantum percolation problems.
However, estimates of the critical exponent $\nu$ often differ from
that obtained in the Anderson tight binding model.
For example, renormalization group analyses of the
quantum percolation problem gave $\nu =2.1$ \cite{OC,CO},
$\nu=1.9\pm 0.5$\cite{RS}, and $\nu=1.86\pm 0.05$\cite{RBS},
which are considerably larger
than the recent estimate $\nu=1.57\pm 0.02$ using
the Anderson model \cite{SO,SO2},
as analyzed by the transfer matrix method \cite{MK}.

The transfer matrix method, which has been successfully
used in the study of the Anderson tight binding model,
however, cannot be applied to
the quantum percolation problem.
This is because
there is always a finite probability that
all bonds connecting one plane of cross section to the next
are vanishing, so that the system
breaks into disconnected pieces.
To overcome this difficulty, it was proposed\cite{SLG}
that a missing bond be replaced by a bond whose transfer energy is
very small ($\sim 10^{-5}$) compared to the transfer energy between the
connected bonds.
A network of percolating wires was also considered
where the effect of the missing bonds was exactly taken into account
to calculate the two-terminal conductance \cite{AL}.
Though the percolation thresholds determined in the
above studies are consistent with the analytical studies of 
quantum percolation by the Pade approximation \cite{SAH,MAH0},
the above numerical study of conductance and
an analytical one\cite{chang}
suggest low values of $\nu$ (0.73 and 0.38, respectively)
compared to that in the Anderson model.

The analysis of energy levels is free from such geometrical problems.
The Thouless number calculated from the energy levels in different
boundary conditions gives a good estimate of
the quantum percolation threshold in a three-dimensional system \cite{KN1,KN2},
but the exponent $\nu$ is $1.95\pm 0.12$, which is larger than the estimate
in the Anderson model.
Thus, the estimates of critical exponent are widely scattered.
Even in 2D where all the states are believed to be Anderson-localized \cite{gang},
some of the above studies indicate that there is a 
localization-delocalization
transition \cite{MAH0,KN1}.

Another method to extract information on the Anderson transition from
eigenvalues
is the finite-size scaling analysis of energy level
spacing \cite{SSSLS,OO,HS_3DO,Evangelou_3DO,ZK,BSZK,SZ,KOSO}.
Berkovits and Avishai\cite{BA} applied this method and
obtained reliable values of $\pq$ and 
$\nu$ in the quantum percolation problem.
The obtained value of the exponent $\nu =1.35\pm 0.1$ for
the quantum percolation
problem
coincides with that estimated in the Anderson model
by the energy level statistics \cite{ZK}, $\nu =1.45\pm 0.1$, though
it is still slightly small compared to that estimated by
the transfer matrix method for the Anderson tight binding model.

One of the most striking features of the Anderson transition is that 
it is classified onto a few universality classes according to the symmetry 
under the
operation of time reversal.
Therefore,
to study the quantum percolation problem in relation to the
Anderson transition, the sensitivity of $\pq$ and $\nu$
to the breaking of the time reversal symmetry (TRS) is
a very important subject that needs to be investigated.

Here we study the energy level statistics of the
quantum percolation model and discuss
the effect of the breaking of TRS.
To be specific, we restrict ourselves to the bond percolation
problems.
A peculiar form of the universal function of the level
spacing distribution at the critical point,
which is related to the sensitivity
to the boundary conditions, is also reported.

To describe the 3D quantum bond percolation model, we consider the 
following
simple Hamiltonian,
\begin{equation}
H=\sum_{\langle ij \rangle}(t_{ij}a_i^{\dagger}a_j+{\rm h.c}),
\end{equation}
where $\langle ij \rangle$ denotes the nearest neighbors.
The transfer 
integral is defined as
\begin{equation}
t_{ij}=\cases{
        \exp(-2\pi{\rm i}\phi_{ij})   & ({\rm for connected bond}) \cr
        0 & ({\rm for disconnected bond})
},
\end{equation}
where the energy unit is the absolute value of the transfer energy
between connected bonds.
Bonds are
randomly connected with probabilities $p$. $\phi_{ij}$ is the Peierls phase
due to magnetic fields.
When all the Peierls phases are set to 0, 
the Hamiltonian is time reversal symmetric, and
we call it the time reversal symmetric (TRS) model.
When the phases are not vanishing, the Hamiltonian is generally not
time reversal symmetric.
We set
$-\pi<\phi_{ij}<\pi$ randomly, and call this situation
non-TRS model hereafter.
The underlying lattice is a 
three-dimensional cube of length $L$ with periodic boundary conditions
imposed.

For each realization of connected bonds,
we first identify the maximally connected percolating cluster,
and then we diagonalize the Hamiltonian corresponding to this cluster
by Lanczos method.
The  calculation is performed for $N$ different realizations of random bond
configurations, where
$N=580, 300, 175$ and 110 for sizes
$L^3=12^3,15^3,18^3$ and $21^3$, respectively. 
These parameters are chosen so that the number of eigenvalues for each
system size exceeds $10^6$.

We analyze the eigenenergies in the region
where the density of states is smooth (Fig. 1).
In the actual simulation, we take
$|E|=0.2\sim 0.8$. 
Energy spectra are then unfolded \cite{mehta,OKSOK}, and 
the distribution function $P(s)$ of adjacent level spacings $s$
is calculated, which is normalized as
\beq
\int_0^\infty \del s P(s) =1 \; ,
\quad \int_0^\infty \del s s P(s) =1 .
\eeq
For a sufficiently large size, we expect the Poissonian behavior
$P_{\rm P}(s)=\exp(-s)$ for localized states, while it should be the
Wigner-Dyson type for extended states\cite{DM,mehta} where
$P_{\rm Wig}(s)\propto s^\beta \exp(-A_\beta s^2)$ ($\beta=1$ in the 
presence
of TRS and $\beta=2$, otherwise).
Varying the probability $p$ and the size $L$, we extract the information
of the quantum percolation transition.

Plots of $P(s)$ as a function of the bond occupation probability for 
the TRS model and the non-TRS model ($L=21$) are displayed in Fig.
~\ref{lsd}. 
It is clearly seen that the expected
crossover from a Poissonian behavior to that of Wigner-Dyson type
is manifested as we increase the value of $p$. 

In order to obtain the critical value of the 
probability $\pq$ and the critical exponent for the divergence
of the localization length $\nu$, we define
$I(s)$ and $\Lambda(p,L)$ as\cite{HS_3DO}
\begin{equation}
I(s)=\int_0^sP(s'){\rm d}s' ,
\end{equation}
and
\begin{equation}
\Lambda(p,L)= \frac{\int_0^{s_0} I(s){\rm d}s-\int_0^{s_0}
I_{\rm P}(s){\rm d}s}
{\int_0^{s_0} I_{{\rm Wig}}(s){\rm d}s-\int_0^{s_0} I_{\rm P}(s){\rm d}s}.
\end{equation}
Here we set $s_0=1.2$.
We also use other quantities
to characterize the
critical behavior of level statistics,\cite{SZ} and have confirmed that the
results of the transition points and the critical exponents
are almost the same.

Denoting $\xi(p)$ as the localization length or the correlation length
which diverges as
\beq
\xi(p)\sim\frac{1}{|p-\pq|^\nu} ,
\eeq
this function is 
expected to behave as 
\begin{eqnarray}
\Lambda(p,L) &=& f[L/\xi(p)] \nonumber\\
                     &=& a_0+a_1[L(p-p_q)^{\nu}]^{1/\nu}+ \nonumber \\
                     &\ &a_2([L(p-p_q)^{\nu}]^{1/\nu})^2+\cdots
\label{eqn_fit}
\end{eqnarray}
near the critical probability $\pq$.

In Fig.~\ref{lam1}, we show
$\Lambda(p,L)$ as a function of $p$ for different
sample sizes $L=12,15, 18$ and 21.
All curves cross at a single point.
By fitting $\Lambda(p,L)$ to eq.(\ref{eqn_fit}),
we can estimate the critical exponent as well as $\pq$.

In Table I, we summarize our results of the $\chi^2$ fit
to  eq.(\ref{eqn_fit}).
Here, the error bars of raw data as well as the fitting
parameters are estimated using the
bootstrap method \cite{SO2,SO,NUMREP,DAVISON}.
Sizes less than $L=9$ give an unreasonably large
$\chi^2$,
so we have excluded them.

The earlier result by Berkovits and Avishai\cite{BA} for the TRS model
gives $1.35\pm 0.1$, which is smaller than the transfer
matrix result for the Anderson model, $1.57\pm 0.02$ \cite{footnote}.
Here we have considered two possibilities:
(i) the effect of energy levels of small disconected clusters,
and (ii) the corrections to scaling due to small sizes.

Let us begin with the first possibility.
Including the levels of
small disconnected clusters \cite{BA}, in principle, influences
the results, since levels in the isolated clusters are
not correlated with those in the percolating cluster,
which destroys the level correlation. 
However, the results from all the energy levels (TRS(all)) almost
coincide with those extracted from the information of the eigenenergies
only for the states on the percolating cluster  (TRS).
This might be due to the fact that we are observing the region
far from the classical percolation threshold $\pc\approx 0.249$,
and most of the states belong to the percolating clusters.\cite{avishai}
We conclude that the possibility (i) is irrelevant, and
thus discuss next possibility (ii).

The analysis by finite-size scaling of the level statistics
tends to give smaller values\cite{SSSLS,HS_3DO,ZK,BA}
compared to that by the transfer matrix
method combined with finite-size scaling \cite{KM,SO}.
It is possible that such discrepancies arise from the
small size of the system, where corrections to scaling
are not negligible \cite{huckestein,WJL,polyakov,SO2}.
To investigate the corrections to scaling, we omit the data for
smaller system sizes \cite{KKO}.
In the TRS model, excluding smaller system sizes tend to increase $\nu$,
which saturates around $\nu\approx 1.5$, in agreement with the
Anderson model.

Breaking the time reversal symmetry (non-TRS model) decreases the value
of the transition point $\pq$\cite{MAH}.
This can be attributed to the destruction of the
constructive quantum interference among the time-reversed paths,
which supports the idea that this transition is indeed the
Anderson transition.
In the non-TRS model, $\nu$ tends to increase by excluding smaller
system sizes, but the fitting becomes unstable when only the data for
$L=18$ and 21 are used.
We conclude that the value of $\nu$ in the non-TRS model
is consistent with that of the Anderson mdoel
belonging to the unitary universality class\cite{SO} ($\nu=1.43\pm 0.02$).

Now we focus our attention on the level statistics at the critical point
$\pq$.
In Fig. \ref{pcs}, we plot the distribution functions $P(s)$ for
the nearest-neighboring level spacing $s$ at $\pq$ for various
system sizes.
They seem to be independent of the sizes, but are different
from that for the Anderson model with the periodic boundary condition
(p.b.c.)
and from that with the fixed boundary condition (f.b.c.).
Recently, it was reported that the form of $P(s)$ at the critical point
changes with the change of boundary conditions\cite{BMP}.
The critical distribution function $P(s)$ obtained in the present study
lies between the two extreme cases.
This is consistent with the fact that some of the bonds are missing
at the edges of the system in the percolation problem,
so even if we adopt the periodic boundary conditions, the
model is effectively closer to the one with f.b.c.
Here, we have plotted the results for the non-TRS model, but
the same behavior is also observed in the TRS model.

In conclusion, we have analyzed the three-dimensional
percolation problems by the finite-size scaling
analysis for the energy levels.
Breaking the time reversal symmetry lowers the quantum percolation
threshold.
Both the time reversal symmetric and non-symmetric models
exhibit the critical exponent $\nu$ consistent with the results of
Anderson tight binding model.
At the critical point, the energy level spacing distribution, which
characterizes the level statistics, becomes size-independent,
which differs from the one in the Anderson model with p.b.c. or f.b.c.

At present, we cannot clearly show that the TRS model and the
non-TRS model belong to different universality classes.
It took large-scale numerical effort to distinguish
one universality class from the other even for the Anderson
model \cite{SO}, since highly accurate data of less than $0.2\%$
relative error are required.
This high accuracy is very difficult to achieve by level statistics
for the quantum percolation problem.
Much more numerical effort is necessary for this purpose.

The authors would like to thank Professors Y. Avishai and
K. Slevin for useful comments.

\vfill\eject

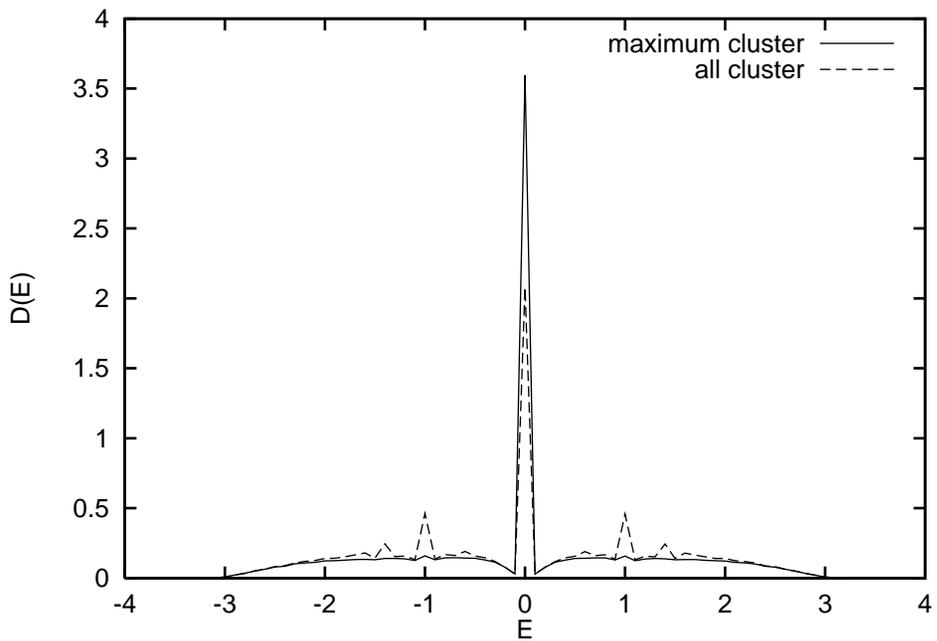
\begin{figure}
\caption{Normalized density of states of the quantum percolation problem 
for
the non-TRS model.
The solid line is for the maximally connected cluster, while the
dashed line corresponds to all the energy levels including
small clusters.}
\label{dos}
\end{figure}

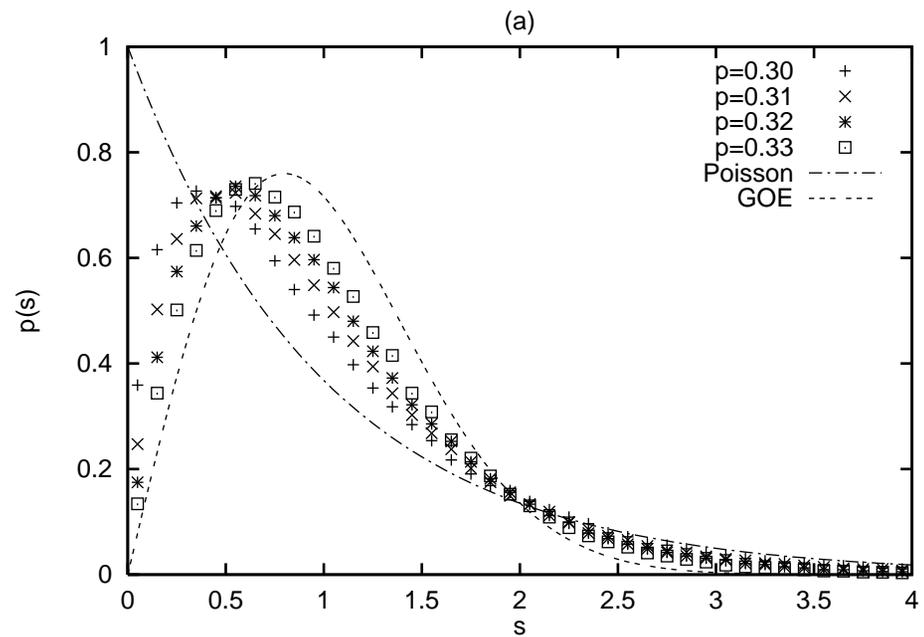
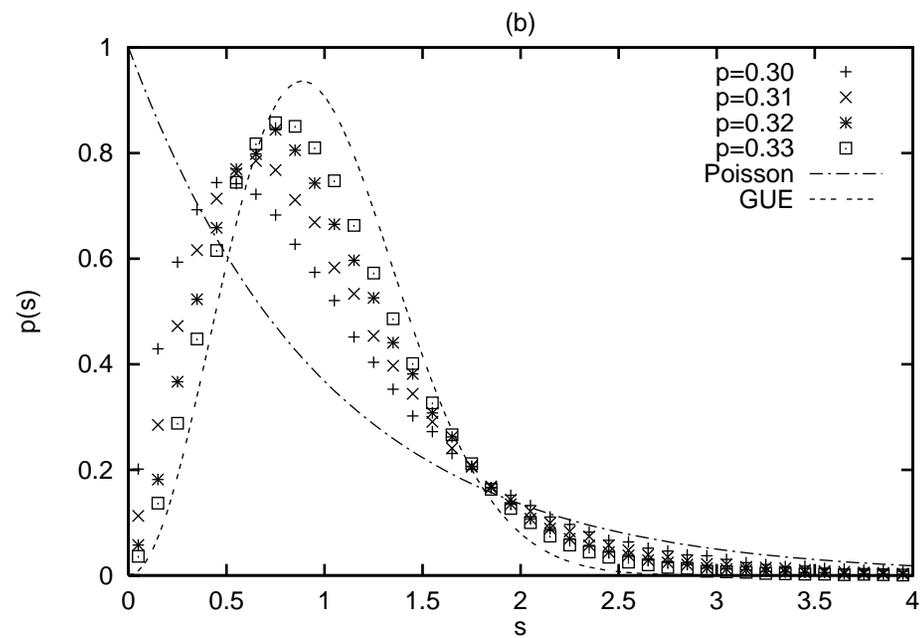
\begin{figure}
\caption{Level spacing distribution $P(s)$ for the TRS model (a) and 
the non-TRS model(b)
for $L=21$. With an increase in $p$, the transition from a Poisson 
distribution 
to that of  Gaussian orthogonal ensemble is seen in Fig. 2(a), while
the transition from a Poisson distribution
to that of Gaussian unitary ensemble is seen in Fig. 2(b). }
\label{lsd}
\end{figure}

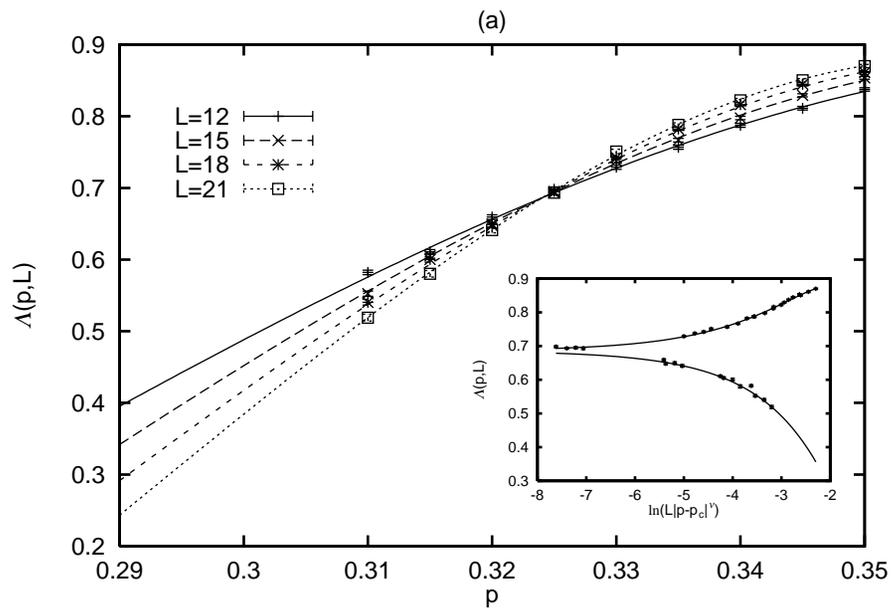
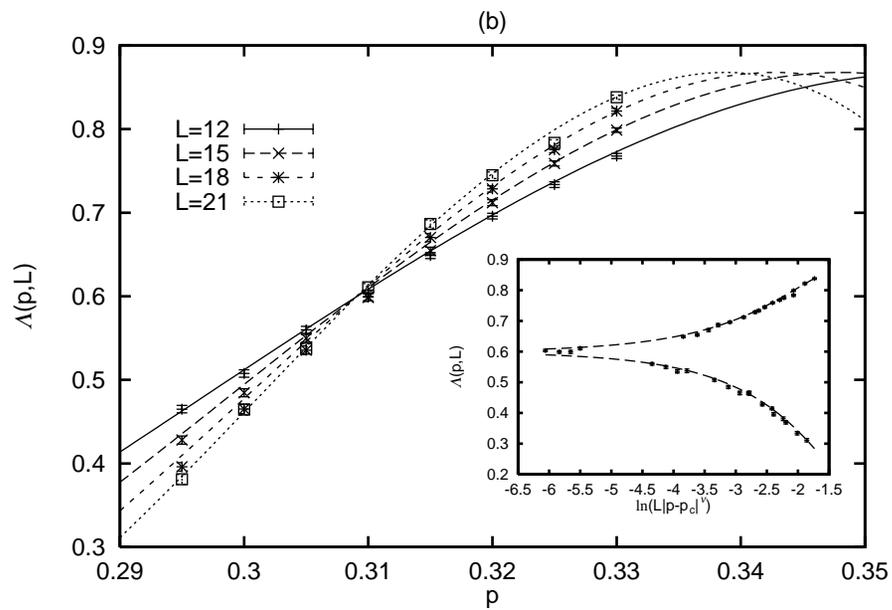
\begin{figure}
\caption{$\Lambda(p,L)$ as a function of $p$ for 
different sample sizes for the TRS model (a)  and the non-TRS model(b) 
obtained from the energy of the percolating cluster.
Energy range is 
$|E|=0.2\sim 0.8$. At $\pq$, all curves cross at a point where
$\Lambda(p,L)$ becomes scale-independent.
Bars around the data points indicate error bars estimated by
the bootstrap method, and the curves are
the results of the fit to eq.(\ref{eqn_fit}).
Insets are  the scaling plots.}
\label{lam1}
\end{figure}

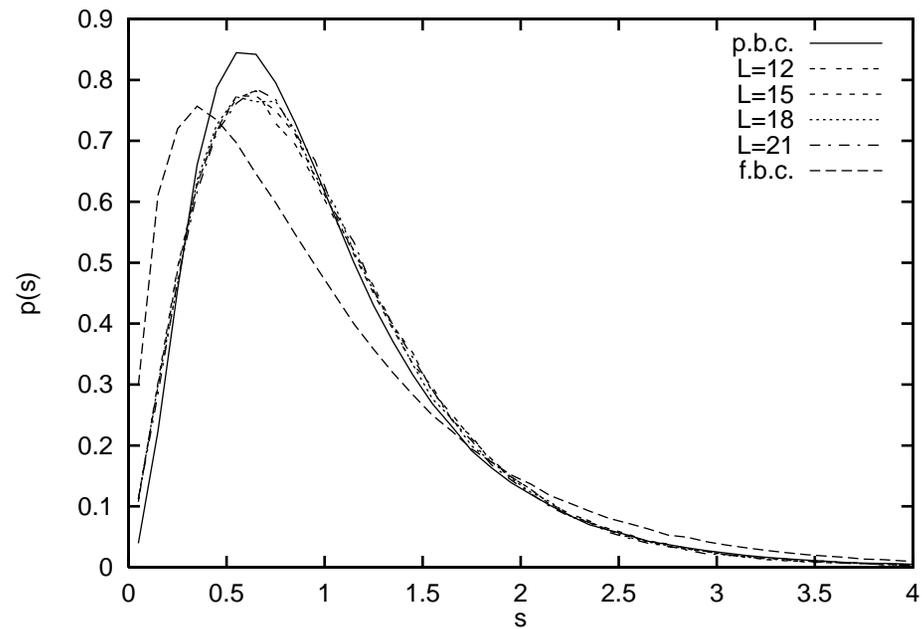
\begin{figure}
\caption{Critical level spacing distribution for
the non-TRS model, which shows intermediate behavior between
that of Anderson model with the p.b.c. (solid line) and those with
f.b.c. (dashed line).}
\label{pcs}
\end{figure}


\begin{table}[hc]
\caption{The $\chi^2$ fitting of critical point $p_q$ and critical 
exponent $\nu$ with their standard deviations.
The values of the TRS model and the non-TRS model are only for
percolating cluster,  while TRS (all)
is obtained from all energy levels. 
TRS (all,BA) is the result reported
by reported Berkovits and Avishai [30].}
\label{cp}
\begin{tabular}{@{\hspace{\tabcolsep}\extracolsep{\fill}}llll} \hline
system    & size                &    $\pq$           & $\nu$  \\ \hline
TRS (all,BA) & $L=7,9,11,13,15$ & $0.33\pm .01$      & $1.35\pm .10$ \\
TRS (all) & $L=12,15,18,21$     & $0.321\pm .001$    & $1.48\pm .10$ \\
TRS       & $L=12,15,18,21$     & $0.324\pm .001$    & $1.46\pm .09$ \\
TRS       & $L=15,18,21$        & $0.323\pm .001$    & $1.55\pm .15$ \\
TRS       & $L=18,21$           & $0.327\pm .002$    & $1.50\pm .28$ \\ 
\hline
non-TRS   & $L=12,15,18,21$     & $0.309\pm .001$    & $1.24\pm .08$ \\
non-TRS   & $L=15,18,21$        & $0.308\pm .001$    & $1.44\pm .16$ \\ 
\hline
\end{tabular}
\end{table}


\begin{thebibliography}{99}
\bibitem{stauffer} D. Stauffer and A. Aharony:
{\it Introduction to Percolation Theory} (Taylor and Francis, London,
1994).
\bibitem{shapiro} B. Shapiro: in
{\it Percolation Structures and Process}, ed. G. Deutscher {\it et al.}
Ann. Isr. Phys. Soc. (1983) 367.
\bibitem{anderson} P.W. Anderson: Phys. Rev. {\bf 109} (1958) 1492 .
\bibitem{gang} E. Abrahams, P.W. Anderson, D.C. Liciardello and
T. V. Ramakrishnan: Phys. Rev. Lett. {\bf 42} (1979) 673.
\bibitem{kawabata} A. Kawabata: Prog. Theor. Phys. Suppl. {\bf 84} 
(1985)  16.
\bibitem{LR} P. A. Lee and T. V. Ramakrishnan: Revs. Mod. Phys.
{\bf 57} (1985) 287 .
\bibitem{KM} B. Kramer and A. Mackinnon: Rep. Prog. Phys. {\bf 56}
 (1993) 1469.
\bibitem{OC} T. Odagaki and K.C. Chang: Phys. Rev. B{\bf 30} (1984) 1612.
\bibitem{CO} K.C. Chang and T. Odagaki: J. Phys. A {\bf 20} (1987) L1027.
\bibitem{RS} L. Root and J.L. Skinner: Phys. Rev. B{\bf 33} (1986) 7738.
\bibitem{RBS} L. Root, J.D. Bauer and J.L. Skinner:
Phys. Rev. B {\bf 37} (1988) 5518.
\bibitem{SO} K. Slevin and T. Ohtsuki: Phys. Rev. Lett. {\bf 78}  (1997) 
4083.
\bibitem{SO2} K. Slevin and T. Ohtsuki:
Phys. Rev. Lett. {\bf 82}  (1999)382.
\bibitem{MK} A. MacKinnon and B. Kramer: Phys. Rev. Lett. {\bf 47} 
(1981)1546;
Z. Phys. B {\bf 53} (1983)1.
\bibitem{SLG}  C.M. Soukoulis, Q. Li and G.S. Grest: Phys. Rev. B {\bf 45}
(1992) 7724.
\bibitem{AL}  Y. Avishai and J.M. Luck: Phys. Rev. B {\bf 45}
(1992) 1074.
\bibitem{SAH}  Y. Shapir, A. Aharony and A.B. Harris: Phys. Rev. Lett. {\bf
 49}
(1982) 486.
\bibitem{MAH0}  Y. Meir, A. Aharony and A.B. Harris: Europhys. Lett. {\bf 
10}
(1989) 275.
\bibitem{chang} I. Chang {\it et al.}: Phys. Rev. Lett. {\bf 74} (1995) 
2094.
\bibitem{KN1}  Th. Koslowski and W. von Niessen: Phys. Rev. B {\bf 42}
(1990) 10342.
\bibitem{KN2}  Th. Koslowski and W. von Niessen: Phys. Rev. B {\bf 
44}(1991) 9926.
\bibitem{SSSLS} B.I. Shklovskii, B. Shapiro, B.R. Sears, P. Lambrianides
and H.B. Shore: Phys. Rev. {\bf B47} (1993) 11487.
\bibitem{OO} Y. Ono and T. Ohtsuki: J. Phys. Soc. Jpn. {\bf 62}  (1993)
3813.
\bibitem{HS_3DO} E. Hofstetter and M. Schreiber: Phys. Rev. {\bf B48}
(1993) 16979 ; {\bf B49} (1994) 14726.
\bibitem{Evangelou_3DO} S.N. Evangelou: Phys. Rev. {\bf B49} (1994) 16805.
\bibitem{ZK} I. Kh. Zharekeshev and B. Kramer: 
 Jpn. J. Appl. Phys. {\bf 34} (1995)4361; 
 Phys. Rev. {\bf B51} (1995) 17239;
 Phys. Rev. Lett {\bf 79} (1997) 717.
\bibitem{BSZK} M. Batsch, L. Schweitzer,  I. Kh. Zharekeshev and B. Kramer:
Phys. Rev. Lett. {\bf 77} (1996) 1552 .
\bibitem{SZ} L. Schweitzer and  I. Kh. Zharekeshev:
J. Phys. Cond. Matt. {\bf 9} (1997) L377.
\bibitem{KOSO}T. Kawarabayashi, T. Ohtsuki, K. Slevin and Y. Ono:
Phys. Rev. Lett. {\bf 77} (1996) 3593.
\bibitem{BA} R. Berkovits and Y. Avishai: Phys. Rev. B {\bf 53}
(1996)16125 .
\bibitem{mehta} M.L. Mehta: {\it Random Matrices} 2nd Edition 
(Academic Press, San Diego, 1991). 
\bibitem{OKSOK}Y. Ono, H. Kuwano, K. Slevin, T. Ohtsuki and B. Kramer:
J. Phys. Soc. Jpn. {\bf 62} (1993) 2762.
\bibitem{DM} F.J. Dyson and M.L. Mehta: J. Math. Phys. {\bf 4} (1963) 701 .
\bibitem{NUMREP} Ch. 15, {\it Numerical Recipes in Fortran},
W. Press, B. Flannery and S. Teukolsky, (Cambridge Univ. Press, Cambridge, 
1992).
\bibitem{DAVISON} Chs. 4-5 {\it Bootstrap Methods and their Application},
A. C. Davison and D. V. Hinkley, (Cambridge Univ. Press, Cambridge, 1997).
\bibitem{footnote}
Most literature use a standard deviation for the error bar of
the exponent, but twice the standard deviation (95$\%$ confidence
interval) was used in Slevin and Ohtsuki.\cite{SO,SO2}
In order to avoid confusion, we cite the values by them with
the error bar of single standard deviation.
\bibitem{avishai} Y. Avishai: private communication.
\bibitem{huckestein} B. Huckestein: Phys. Rev. Lett. {\bf 72} (1994) 1080.
\bibitem{WJL} Z. Wang, B. Jovanovic and D.-H. Lee: Phys. Rev. Lett.
{\bf 77} (1996) 4426 .
\bibitem{polyakov} D.G. Polyakov: Phys. Rev. Lett. {\bf 81} (1998) 4696.
\bibitem{KKO} T. Kawarabayashi, B. Kramer and T. Ohtsuki :
Phys. Rev. {\bf B57} (1998) 11842.
\bibitem{MAH} Y. Meir, A. Aharony and A. Harris: Phys. Rev. Lett. {\bf 56}
(1986) 976. In this reference, uniform magnetic fields are applied and
a cyclotron motion is present, while in our
random phase model, such motion is absent and the shift of $\pq$ is
purely from the breaking of TRS.
\bibitem{BMP} D. Braun, G. Montambaux and M. Pascaud:
Phys. Rev. Lett. {\bf 81} (1998) 1062.

\end{thebibliography}
\end{document}